\documentclass[fleqn,usenatbib]{mnras}

\usepackage{newtxtext,newtxmath}

\usepackage[T1]{fontenc}

\DeclareRobustCommand{\VAN}[3]{#2}
\let\VANthebibliography\thebibliography
\def\thebibliography{\DeclareRobustCommand{\VAN}[3]{##3}\VANthebibliography}

\usepackage{graphicx}	
\usepackage{amsmath}	

\title[Detectability of Satellite Planes]{Detectability of Satellite Planes in Mock Observations of Isolated $L_*$ Galaxies}

\author[E.~Crosby et al.]{Ethan Crosby$^{1}$\thanks{E-mail: Ethan.Crosby@anu.edu.au},
Marcel S. Pawlowski$^{2}$,
Oliver M\"uller$^{3,4,5}$,
Helmut Jerjen$^{1}$
\\
$^{1}$Research School of Astronomy and Astrophysics, Australian National University, Canberra, ACT 2611, Australia\\
$^{2}$Leibniz-Institut f\"ur Astrophysik Potsdam, An der Sternwarte 16, D-14482 Potsdam, Germany\\
$^{3}$Institute of Physics, Laboratory of Astrophysics, Ecole Polytechnique Fédérale de Lausanne (EPFL), 1290 Sauverny, Switzerland\\
$^{4}$Visiting Fellow, Clare Hall, University of Cambridge, Cambridge, UK\\
$^{5}$Institute of Astronomy, Madingley Rd, Cambridge CB3 0HA, UK\\
}

\date{Accepted XXX. Received YYY; in original form ZZZ}

\pubyear{2025}

\begin{document}
\label{firstpage}
\pagerange{\pageref{firstpage}--\pageref{lastpage}}
\maketitle

\begin{abstract}
The existence and prevalence of planar, co-rotating distributions of satellite galaxies around $L_*$ host galaxies in the local universe remains a subject of ongoing debate. Despite numerous observational efforts over the past decade, a statistically robust sample of "satellite planes" across the diversity of host galaxy environments is lacking. To guide future observing strategies, we construct a controlled suite of mock observations of on-sky positions and line-of-sight (LOS) velocities of isolated $L_*$ host galaxies and their satellite systems, based on samples drawn from the \textsc{Illustris} \textsc{TNG100-1} cosmological simulation to build a statistical sample. In these mock systems, satellite planes are defined by three key parameters: the number of satellites ($N_{\mathrm{sat}}$), the fraction residing in a thin co-rotating plane ($f_{p}$), and the orientation angle relative to the observer ($\theta_{\mathrm{rot}}$). We evaluate the sensitivity of three observational metrics, $N_{\mathrm{cor}}$ (number of co-rotating satellites), $b/a$ (projected flattening of the satellite distribution), and $v_\mathrm{los}$ (mean absolute LOS velocity), to the presence of such planes. Our results show that detection rates are strongly dependent on $\theta_{\mathrm{rot}}$ and $N_{\mathrm{sat}}$. Satellite planes that are viewed nearly edge-on or face-on, are the most readily detected. In contrast, intermediate orientations and systems with fewer satellites yield low detection success rates. Generally, only satellite planes with $N_{\mathrm{sat}}>20$ have high chances of being detected. These findings provide a practical framework for prioritising observational targets and designing future surveys aimed at detecting and characterising satellite planes.

\end{abstract}



\section{Introduction}
For many decades, the classical satellite dwarf galaxies of the Milky Way, roughly those dwarf galaxies with $M_{V}<-8.5$, have been known to be aligned in a thin disk-like structure \citep{LyndenBell1976}. Conversations beginning in the 2000s highlighted how this thin disk, or 'satellite plane', was not commonly reproduced by cosmological models \citep{Kroupa_2005,Metz_2009_a}, thus appearing to be an outlier of $\Lambda$CDM cosmology. Immediately, a continued back and forth discussion emerged which provided hypotheses for physical mechanisms that could produce a satellite plane, including the simultaneous infall of dwarf galaxy groups \citep{Li2008,DOnghia2008,Metz2009_B,Jlio2024}, tidal dwarf galaxies rotating in thin planes following galactic interactions \citep{Kroupa_2005,Metz2007_b,Pawlowski2012_a,Kroupa_2012,Hammer_2013} or hierarchical accretion along flattened or thin cosmic filaments \citep{Libeskind2010,Lovell2011}. However, this later hypothesis is unlikely the original of thin planes \citep{Pawlowski2012_b}. A number of papers highlighted possible biases influencing the results, including the temporal stability of these planes \citep{Bahl_2014,Gillet_2015,Buck_2016,Shao_2019,Taibi2024}, the 'look-elsewhere' effect \citep{Cautun2015}, the influence of broader cosmological structures, namely the 'Local Sheet' \cite{Libeskind_2019} and the dynamics of the Large Magellanic Cloud \citep{Samuel2021,GaravitoCamargo2021}, which is a disputed topic \citep{Pawlowski2022}. The discussion continues, with the most recent results incorporating full spatial and kinematic vectors of the Milky Way to showing this extraordinary plane remains in contention with $\Lambda$CDM simulations \citep{Pawlowski2015,Pawlowski_2019,Pawlowski_2021_a}, and similar satellite planes being identified around M31 \citep{Ibata_2013,Conn2013,Pawlowski_2021_b}, Centaurus A \citep{Muller2018, Mueller_2019b, Muller2021} and NGC4490/NGC4485 \citep{Karachentsev2024,Pawlowski_2024}.

These systems collectively lie within the Local Sheet, which as identified above is a possible source of bias, given there appears to be a global preferential axis of satellite galaxy orientations \citep{Libeskind_2019} and that the Local Sheet itself appears to be a rare outlier in the context of $\Lambda$CDM simulations \citep{Neuzil2020,AragonCalvo2022}. This is partly motivating numerous efforts to identify more satellite dwarf galaxies, particularly those beyond the Local Sheet, further than $\sim7-8\,$Mpc \citep{Tully2008,McCall2014,Anand2019}. Recent surveys have collectively built an extensive list of dwarf satellite candidate galaxies found in host galaxy environments on the edge of the Local Volume (a $\sim10\,$Mpc sphere) \citep{Javanmardi_2016,Muller_2017,Muller_2018_c,Byun_2020,Danieli2020,Muller_2020,Heesters2021,MartnezDelgado2021,Carlsten_2022,Crosby_2023_a,Crosby_2023_b}. 

However, these optical images alone are not sufficient to robustly quantify the presence or absence of a satellite plane. Some degree of anisotropy is expected for simulated $\Lambda$CDM satellite systems \citep{Libeskind2005,Zentner2005} such that 2D on-sky coordinates alone do not commonly reveal anisotropy that stands in contention with $\Lambda$CDM simulations \citep{Crosby_2023_a,Crosby_2023_b}. Well defined four dimensional (4D) satellite planes are classified with at least the addition of distance measurements (the third spatial coordinate) and a line-of-sight (LOS) velocity. For example, the satellite planes of Centaurus A and M31 are largely 4D satellite planes \citep{Ibata_2013,Muller2021}. With the addition of proper motions, the satellite plane of the Milky Way is defined in all six dimensions \citep{Pawlowski_2021_a}.

The distance measurements used to define a satellite plane must have uncertainty no greater than $\sim\pm100\,$kpc in most cases, since that is the typical scale of a satellite plane system around Milky Way type hosts ($M_{*}\sim10^{10.5}\,M_{\odot}$). The most precise distance measurement method for the predominately early-type dwarf satellites galaxies is the Tip of the Red Giant Branch (TRGB) method. This requires the resolution of individual RGB stars, which places constraints on the FWHM of point sources in the image. As objects become more distant, the angular size of FWHM must decrease in order to resolve individual stars. Historically, ground based telescopes are able to employ this method for galaxies out to $\sim3\,\mathrm{Mpc}$ which then become limited by the FWHM of points sources beyond this distance. The Hubble Space Telescope (HST), is able to employ this analysis out to $\sim10\,\mathrm{Mpc}$ for most observations, where it becomes limited by the apparent magnitude of stars \citep{Jacobs2009}. But even HST can only reliably achieve the desired level of precision, $\sim\pm100\,\mathrm{kpc}$, within the $\sim7\,\mathrm{Mpc}$ sphere (as in the best-case scenario from the Extragalactic Distance Database \citep{Jacobs2009}), so the need to expand the search to satellite planes beyond the $\sim7-8\,$Mpc mark presents a serious challenge; that precise enough distance measurements will be very difficult to acquire.

However, LOS velocities acquired through optical spectroscopy is achievable with spectrographs in ground-based telescopes out to distances of around $100\,\mathrm{Mpc}$. In the near term, any attempt to define a satellite plane beyond $\sim7-8\,$Mpc will be limited to 2D on-sky coordinates plus the LOS velocity. Methods based on these more limited coordinates are being used to characterise satellite planes of distant systems such as the M104 \citep{Crosby2024}, NGC2750 \citep{Paudel2021} and NGC4490/NGC4485 \citep{Karachentsev2024,Pawlowski_2024}.

However, even in this hotly debated topic, receiving the large amount of observing time required on 8m-class telescopes with optimal spectroscopic instruments, which are among the most oversubscribed instruments in astronomy, is challenging. Although some recent large-scale surveys have begun this task \citep{Heesters2021}, only the most remarkable systems receive the complete and valuable targeted follow-up required to properly assess the presence of satellite planes as in \cite{Muller2025}. This piecemeal reporting of individual satellite galaxy systems favors the presentation of only the most outstanding and well-defined satellite planes, which could create a selection bias that leads to an incorrect perception of satellite plane ubiquity. In fact, statistically and robustly demonstrating that a system contains no satellite planes would be remarkable on its own, but few observing programs have considered this as a primary goal.

This paper seeks to present a suite of simulations and statistical measures that can guide and enhance the proposals of prospective future observing programs by quantifying the requirements needed to confirm the presence or absence of planar alignments of satellite galaxies.

\section{Method} \label{sec:method}

In this section, we construct a controlled suite of mock observations of satellite systems with known satellite plane configurations. Our goal is to generate a statistically robust sample of synthetic observations that can probe the relationship between measurable quantities accessible to observers and the underlying structural properties of satellite planes. In the absence of being able to measure the complete position and velocity vectors of satellites, for reasons discussed in the introduction, we focus on exploring how readily accessible observational quantities from more limited data—just on-sky positions and line of sight velocities—can be used to infer the underlying structure of satellite systems.

To perform this assessment we require a suite of mock-observations. It will consist of observations of spatially isolated $L_*$ galaxies and their satellite systems. To construct it, we adopt the following simplifying assumptions:

\begin{enumerate}
    \item Satellite plane membership is determined solely by the position and velocity vectors of satellites, within an otherwise simplified host galaxy environment. We ignore any potential correlations between satellite plane membership and satellite luminosity, stellar age, morphology, or other secondary properties. This is a reasonable assumption based on trends in the Milky Way and M31 systems \citep{Collins_2015,Taibi_2024}
    \item The only difference between satellites within and outside a plane lies in the orientation of their position and velocity vectors. Satellites in a plane are assumed to have the same radial and velocity magnitude distributions as those outside of it.
\end{enumerate}

These assumptions allow us to utilise cosmological $\Lambda$CDM simulations as the foundation for our suite, to reproduce realistic satellite radial distributions and velocity components that are compatible with comparisons to $\Lambda$CDM simulations. 

\subsection{Obtaining realistic radial distributions and velocity components from $\Lambda$CDM}

Accordingly, we select the Illustris TNG100-1 simulation \citep{Springel_2017, Nelson_2017, Naiman_2018, Marinacci_2018, TNG_MAIN, Nelson_2019} as our reference $\Lambda$CDM framework. TNG100-1 is a gravo-magnetohydrodynamical cosmological simulation with a box size of $106.5\,\mathrm{Mpc}$, a baryonic mass resolution of $1.4\times10^{6}\,M_{\odot}$, and a dark matter mass resolution of $7.5\times10^{6}\,M_{\odot}$. These limits correspond to a rough absolute magnitude floor of $M_g \sim -8$ to $-10$, comparable to the detection thresholds of deep optical dwarf galaxy surveys beyond $\sim7$ Mpc \citep{Javanmardi_2016, Heesters2021, Danieli2020, Byun_2020, Carlsten_2022, Crosby_2023_a, Crosby_2023_b}. The following step-by-step process is used to extract environments that resemble realistic observations from the sub-halo and Friends-of-Friends catalogs alone.

\begin{enumerate}
    \item We begin by retrieving all Friends-of-Friends (FoF) groups and their associated subhalos at redshift $z=0$.
    \item We select FoF groups with virial masses in the range $5 \times 10^{10} < M_{200} < 1 \times 10^{13}\,M_{\odot}$, along with their subhalos.
    \item We exclude all dark matter deficient objects of 'non-cosmological origin'; those with the 'SubhaloFlag' flag set to 0 (the implications of this is discussed in Section \ref{sec:discussion}).
    \item We retain only subhalos with absolute magnitudes $M_g < -9$, consistent with the luminosity limit of our intended observational comparisons.
    \item We identify isolated host galaxies as subhalos with stellar mass greater than $10^{10}\,M_{\odot}$ that have no other such galaxy within a $700\,\mathrm{kpc}$ radius.
    \item Finally, we select only those isolated host systems that contain at least 10 satellites within their virial radius ($R_{\mathrm{vir}}$).
\end{enumerate}

This selection yields 920 isolated host systems, encompassing a total of 19,833 satellite galaxies. To summarise, an 'isolated host galaxy' is any sub-halo with stellar mass greater than $10^{10}\,M_{\odot}$, with no other halo with that mass within a $700\,\mathrm{kpc}$ radius, and at least 10 other sub-halos within $R_{\mathrm{vir}}$. The 'satellite galaxies' of these hosts are all sub-halos within $R_{\mathrm{vir}}$ of these hosts with $M_g<-9$, that are of cosmological origin. $R_{\mathrm{vir}}$ is calculated based on the total host mass (dark matter plus stellar plus gas mass of the sub-halo, $M_{\mathrm{tot}}=M_{\mathrm{DM}}+M_{\mathrm{*}}+M_{\mathrm{gas}}$), using the following equation:
\begin{equation}
	R_{\mathrm{vir}} = \sqrt[3]{\frac{3M_{\mathrm{tot}}}{4\pi\left(200\rho_{\mathrm{crit}}(z)\right)}}
\end{equation}

\begin{figure}
	\centering
	\includegraphics[draft=false,width=8cm]{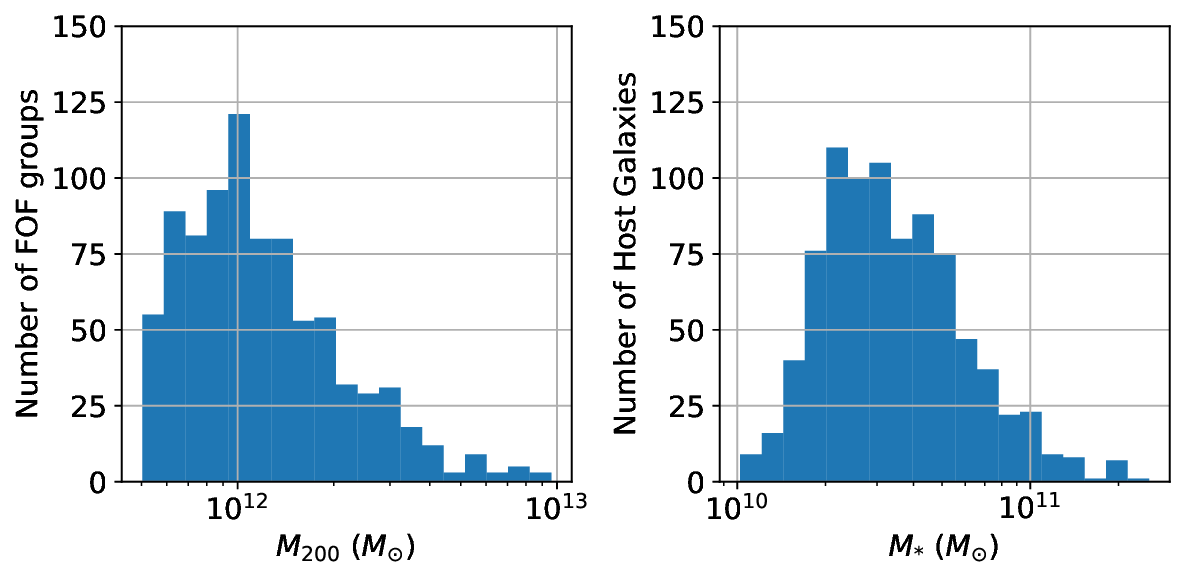}
	\caption{Histograms of the mass distribution of our sample of 920 systems from TNG100-1. In the left histogram, we show the distribution of $M_{200}$, or the virial mass of the FOF groups. In the right histogram, we show $M_*$, or the stellar mass of the host halos that are included in our sample.}
	\label{fig:sample_mass_distributions}
\end{figure}

The majority of the isolated hosts in this sample have stellar masses $M_*\sim1.2\times10^{10}\,M_{\odot}$, which closely resembles Milky Way like systems. The distributions of host $M_*$ and FOF group $M_{200}$ are shown in greater detail in Figure \ref{fig:sample_mass_distributions}.


From this sample, we extract the satellite galaxies and construct a catalog of satellite vectors. Each entry in this catalog includes:



\begin{itemize}
    \item the radial distance to the host,
    \item the radial velocity magnitude,
    \item the tangential velocity magnitude,
\end{itemize}
all defined relative to their host and normalised to the host’s virial radius and virial velocity (i.e. the circular velocity at $R_{\mathrm{vir}}$). This normalisation removes the influence of host mass and allows us to generalise across halos of different sizes. 

\subsection{Constructing Artificial Satellite Systems}

We randomly sample from this catalog of TNG100 selected satellites to construct mock satellite systems. This 'random sampling' involves selecting at random a satellite from our simulated sample, taking its galactocentric distance, and the radial and tangential velocities. This sampling process preserves realistic radial and velocity magnitude distributions (the total velocity magnitude can be calculated from the radial and tangential components), as they are directly drawn from a cosmological simulation that incorporates key aspects of galaxy formation physics. For each mock satellite, the sampled values define the radial distance and the tangential and radial velocity magnitudes. 

The remaining degrees of freedom—specifically the angular coordinates $\theta$ and $\phi$ of the position vector, and the direction in the tangent plane of the sphere for the velocity vector—are set according to whether the satellite is assigned to a co-rotating planar structure or an isotropic distribution.

We define the satellite plane normal vector to initially lie along the positive $z$-axis. Satellites designated as in-plane must have orbital angular momentum vectors (i.e. orbital poles) within $10^\circ$ of this plane normal. These orbital poles are drawn from a uniform distribution over the region in the sphere defined by a 'cap' of a cone subtended by that angle, ensuring both a thin, disk-like spatial structure and coherent co-rotation within the plane. Our choice of $10^\circ$ is motivated by the observed thickness of the Milky Way’s satellite plane \citep{Pawlowski_2019}. Out-of-plane satellites are assigned orbital poles drawn from a uniform distribution over the full unit sphere.

Once an orbital pole is selected for each satellite, we randomly select an orbital angle from a flat distribution in a circle to set its position. The full velocity vector can then be directly calculated. At this stage, we obtain full 3D position and velocity vectors for every satellite.

We control the structure of the co-rotating plane with the following parameters:
\begin{enumerate}
    \item $N_{\mathrm{sat}}:$ The total number of satellites within a single system.
    \item $f_{p}:$ The plane fraction, or the fraction of satellites that are considered 'in-plane'.
    \item $\theta_{\mathrm{rot}}:$ The inclination angle of the plane relative to the line-of-sight, $0^\circ$ means the satellite plane (if present) is viewed edge-on, and $90^\circ$ means it is viewed face-on.
\end{enumerate}

\begin{figure}
	\centering
	\includegraphics[draft=false,width=8cm]{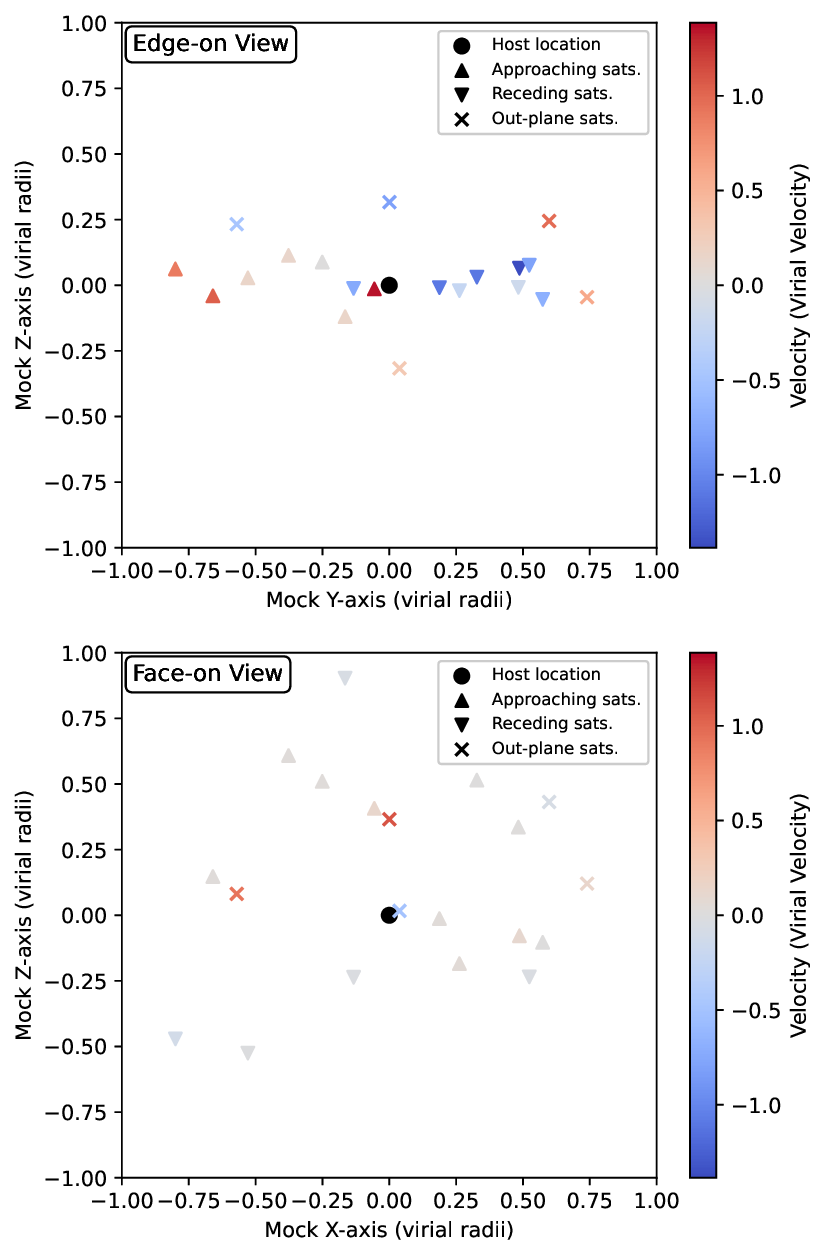}
	\caption{A mock observation of a satellite system containing a satellite plane with $N_{\mathrm{sat}}=20$ and $f_{p}=75\%$. In both graphs, downward triangles represent the positions of in-plane satellites that are blueshifted relative to the host, and upward triangles are redshifted. Crosses represent satellites not in the plane. The large dot indicates the location of the host galaxy. The colour of the symbols indicates the velocity magnitude relative to the virial velocity, and is indicated on the plot within the colour bar. The top panel is an edge-on ($\theta_{\mathrm{rot}}=0^\circ$) view of the satellite plane, bottom panel is a face-on view ($\theta_{\mathrm{rot}}=90^\circ$).}
	\label{fig:sample_views}
\end{figure}

These parameters are used to simulate observations of generated satellite planes. A system of satellites is generated with $N_{\mathrm{sat}}$ members, and then the proportion of those placed in a satellite plane is set by $f_{p}$. Then, a rotation is applied about the $y$-axis by an angle $\theta_{\mathrm{rot}}$ ranging from $0^\circ$ to $90^\circ$. Finally, the system is viewed along the $x$-axis. We perform a mock observation of this configuration, extracting projected 2D satellite positions (on-sky coordinates, or the y-z plane) and their line-of-sight (LOS) velocities. Examples of edge-on and face-on mock observations are shown in Figure \ref{fig:sample_views}.


\section{Results}

\subsection{Satellite Plane Metrics}\label{sec:metrics}

When working solely with on-sky positions and LOS velocities, two key metrics are commonly used to assess the presence and coherence of a satellite plane \citep{Ibata_2013,Muller2018,Paudel2021,Crosby2024,Pawlowski_2024}: (1) the number of co-rotating satellites, $N_{\mathrm{cor}}$, and (2) the spatial flattening of the system, typically quantified via the axis ratio $b/a$ of an ellipse fit to the satellite distribution. Several methods exist for computing these quantities; below, we describe the specific approach used in our analysis.

Our definition of $N_{\mathrm{cor}}$ can be described as the maximum number of co-rotating satellites identified across all possible orientations of a projected rotation axis. To compute this we consider a straight line, interpreted as a candidate orbital axis, drawn through the host galaxy in the plane of the sky. We rotate this axis in $1^\circ$ increments from $0^\circ$ to $180^\circ$, and at each orientation, we compute the number of satellites exhibiting coherent LOS velocity patterns across the axis.

For a given axis orientation, we divide the satellites into two hemispheres (one on each side of the axis) and compute two values:
\begin{itemize}
    \item The sum of the number of blueshifted satellites on the first side and the number of redshifted satellites on the second side.
    \item The reverse configuration: redshifted satellites on the first side and blueshifted on the second.
\end{itemize}
The final reported value of $N_{\mathrm{cor}}$ is the maximum value across all these possible combinations.

To calculate the axis ratio $b/a$, we use a moment-based, weighted ellipse-fitting method applied to the projected satellite distribution in the plane of the sky. The ratio of the semi-minor axis ($b$) to the semi-major axis ($a$) of this ellipse is interpreted as a measure of the system's spatial flattening.

The method begins by constructing the covariance matrix of the satellites' projected $x$–$y$ positions, which encodes the second moments of the distribution. The eigenvalues of this matrix are proportional to the squared lengths of the principal axes of the best-fit ellipse. From these, we compute $b/a$ as the square root of the ratio of the smaller to the larger eigenvalue. However, this approach is known to be highly sensitive to outliers, which can bias the result and reduce its diagnostic power for identifying planar satellite structures.

To mitigate this sensitivity, we apply an iterative reweighting scheme (based on algorithms provided in \citealt{Marazzi1993}) that down-weights outliers based on their distance from the mean of the satellite distribution. Specifically, we use the Mahalanobis distance and is defined as:
\begin{equation} \label{eq:Mahalanobis}
    D_m(\boldsymbol{x}) = \sqrt{\boldsymbol{x}^{T}C^{-1}\boldsymbol{x}}
\end{equation}
Where $\boldsymbol{x}$ is the position vector of a satellite (with the host galaxy assumed to lie at the origin), and $C^{-1}$ is the inverse of the current covariance matrix. We then compute a weight for each satellite using Tukey’s Biweight function:
\begin{equation}
    \psi(D_m) = \left\{ \begin{aligned} 
        \left(1-\left(\frac{D_m}{6c}\right)^{2}\right)^2 \mathrm{for}\ |D_m|<c\\ \\
        0\ \mathrm{for}\ |D_m|\geq c
    \end{aligned} \right.
\end{equation}
Here, $c$ is a tuning parameter that defines the cutoff beyond which satellites are treated as outliers; we set $c = \sqrt{\langle D_m \rangle}$, the square root of the mean Mahalanobis distance across all satellites. Intuitively, $\sqrt{\langle D_m \rangle}$, can be described as the multivariate standard deviation.

This process is performed iteratively. We initialise the weights to unity and compute the covariance matrix of the weighted positions. We then calculate Mahalanobis distances and update the weights using the Biweight function. This is repeated until the weights converge to within a fractional tolerance of $10^{-6}$, which typically occurs within 10 iterations.

\begin{figure}
	\centering
	\includegraphics[draft=false,width=8cm]{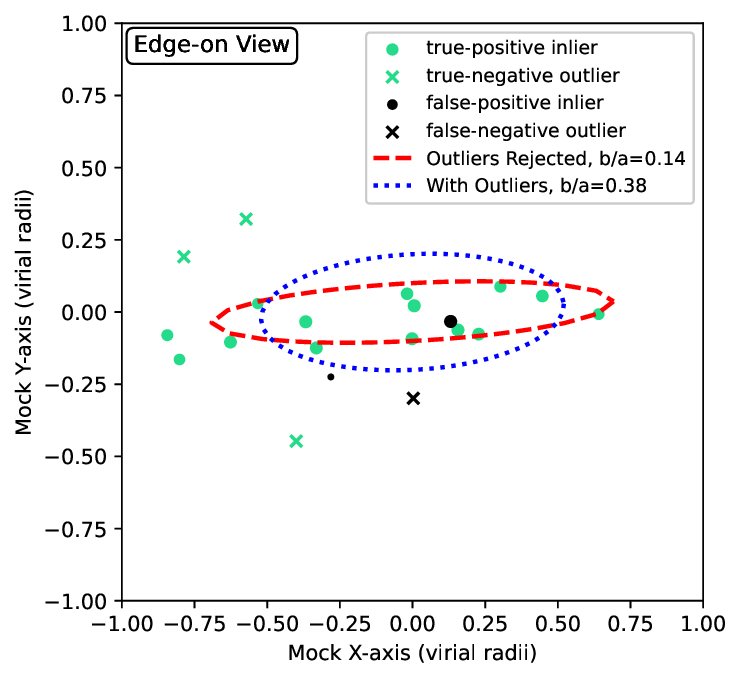}
	\caption{A mock observation of a simulated satellite system with $N_{\mathrm{sat}}=20$, $f_{p}=0.75$ and $\theta_{\mathrm{rot}}=20^\circ$, where the satellites that are determined by the iterative weighting scheme to be outliers (weight is zero) are shown with crosses, and inliers (weight is $>0$) with dots. Whether a satellite was originally generated as in-plane or out-plane is shown by the symbol shape, a circle is an in-plane satellite, and a cross an out-plane satellite. Green symbols have been correctly identified as outliers or inliers, and black symbols incorrectly identified. The size of the green dots is scaled with the weight given to to them. We overlay the figure with an ellipse of the $b/a$ metric from all the satellites (dotted line), and with outliers rejected (dashed line.)}
	\label{fig:outlier_rejection_demonstration}
\end{figure}

The final weighted covariance matrix is used to compute the axis ratio $b/a$, which we report as our measure of spatial flattening. We demonstrate an example of this outlier rejection method in Figure \ref{fig:outlier_rejection_demonstration}, where we visualise the outlier rejection method. We differentiate between the satellites which belong to the generated satellite plane (the 'positive' cases) and which do not (the 'negative' cases). The rejection algorithm then makes its own estimate of the positive and negative cases, which is further split in 'true' cases (a correct prediction) and 'false' cases (an incorrect prediction), such we have four distinct cases, as described in the figure. In this example, rejecting the outliers returns a lower $b/a$, which is more representative of underlying plane distribution. The algorithm performs well in deciding which satellites to include and which to reject in most cases. This method preserves the covariance of the arrangement of points without having to calculate a non-uniform spatial metric when using one-dimensional density based outlier rejection algorithms, as is the case with many functions available in \textsc{scikit-learn} \citep{scikit-learn}, like the Local Outlier Factor (LOF, \citet{Breunig2000}).

This formulation for $N_{\mathrm{cor}}$ and $b/a$ decouples their axes; the rotation axis is not necessarily aligned with the minor axis of the flattening ellipse. From our testing, we found this to be more effective in detecting satellite planes (as described in Section \ref{sec:sim_results}) than coupling the axes of these metrics. However, this can lead to spurious detections of systems where $N_{\mathrm{cor}}$ or $b/a$ suggests a satellite plane is present, but there none in reality. For example, a system with $f_p=0.0$ (no plane) and thus large $b/a$, might also have large $N_{\mathrm{cor}}$ by pure chance. In an observational setting, carefully considering such cases is important and could indicate certain phenomena, such as merger-driven co-rotation (as in the NGC5713/19 system \cite{Jerjen2025}).

The $N_{\mathrm{cor}}$ and $b/a$ metrics are not sensitive to face-on satellite planes (i.e. $\theta_{\mathrm{rot}} = 90^\circ$). In such configurations, the satellite distribution appears nearly circular on the sky and exhibits no measurable co-rotation signal, making it indistinguishable from an isotropic distribution using these metrics alone.

To address this limitation, we introduce a complementary metric based on the magnitude of the LOS velocities. In the case of face-on satellite planes, most of the satellite motion lies in the plane of the sky, resulting in lower-than-expected LOS velocities. This suppression of velocity dispersion along the line of sight can signal the presence of a coherent planar structure, even when it is viewed face-on.

In realistic observational settings, estimating whether LOS velocities are lower than an isotropic distribution requires an independent estimate of the host halo’s virial mass—such as from abundance matching based on the stellar mass of the host galaxy. Otherwise, using a mass estimate based on the motions of the satellites would lead to circular logic when measuring $v_{\mathrm{los}}$.

For our analysis, we define a simple summary statistic, $v_{\mathrm{los}}=\langle|{v_i}|\rangle$, the mean absolute LOS velocity of all satellites in the system. This quantity serves as a proxy for the dynamical temperature of the system along the line of sight and can help identify anomalously cold (i.e., face-on planar) satellite distributions. 

\subsection{Simulation Results}\label{sec:sim_results}

\subsubsection{Basic Distributions of Metrics}

\begin{figure*}
	\centering
	\includegraphics[draft=false,width=18cm]{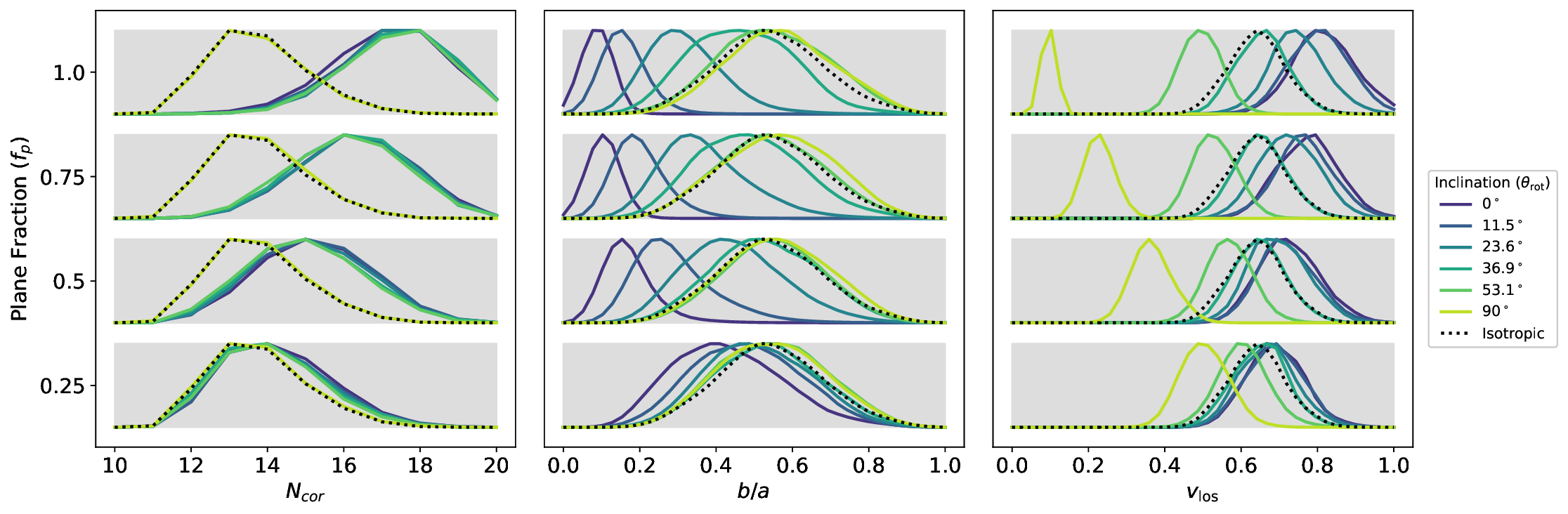}
	\caption{The distributions of the metrics $N_{\mathrm{cor}}$, $b/a$ and $v_{\mathrm{los}}$ across simulations of 20 satellites whose distribution is defined by the parameters $f_p$ and $\theta_{\mathrm{rot}}$. $b/a$ and $v_{\mathrm{los}}$ have been smoothed by a Kernel Density function to approximate the distribution. The value $f_p$ is set by the y-axis, and $\theta_{\mathrm{rot}}$ by the shading of the lines, with a legend in the figure. The dashed line is the reference isotropic case ($f_p=0$) for each metric.}
	\label{fig:simple_KDE}
\end{figure*}

Now we conduct mock observations of our simulated systems consisting of 20 satellites, and compare the distributions of $N_{\mathrm{cor}}$, $b/a$ and $v_{\mathrm{los}}$ across planes that vary as a function of $f_p$ and $\theta_{\mathrm{rot}}$. We select 0.0, 0.25, 0.5, 0.75 and 1.0 as our values of $f_p$, and to best represent the distribution of $\theta_{\mathrm{rot}}$ that will be observed in a realistic scenario, our chosen values are $\mathrm{sin}^{-1}(0)$, $\mathrm{sin}^{-1}(0.2)$, $\mathrm{sin}^{-1}(0.4)$, $\mathrm{sin}^{-1}(0.6)$, $\mathrm{sin}^{-1}(0.8)$ and $\mathrm{sin}^{-1}(1.0)$, which approximately corresponds to 0, 11.54, 23.58, 36.87, 53.13 and 90 degrees respectively. This accounts for the non-uniform distribution of $\theta_{\mathrm{rot}}$ when the plane is randomly oriented as a result of spherical coordinate distributions. For each combination of $f_p$ and $\theta_{\mathrm{rot}}$, we create 10,000 simulated systems by uniformly sampling from the TNG100 sample and measuring $N_{\mathrm{cor}}$, $b/a$ and $v_{\mathrm{los}}$ from mock observations. We use a Kernel Density Estimate to build a smoothed distribution for each metric, and report the results in Figure \ref{fig:simple_KDE}.

Figure \ref{fig:simple_KDE} reveals the following trends. The distribution of $N_{\mathrm{cor}}$ is largely independent of inclination, except for the face-on case ($\theta_{\mathrm{rot}}=90^\circ$). $b/a$ has a significant correlation with inclination, but the width of the distribution means that only nearly edge-on cases ($\theta_{\mathrm{rot}}=0^\circ$) are distinguishable from others. $v_{\mathrm{los}}$ is more or less the opposite, where it is best at distinguishing face-on cases.

\begin{figure*}
	\centering
	\includegraphics[draft=false,width=18cm]{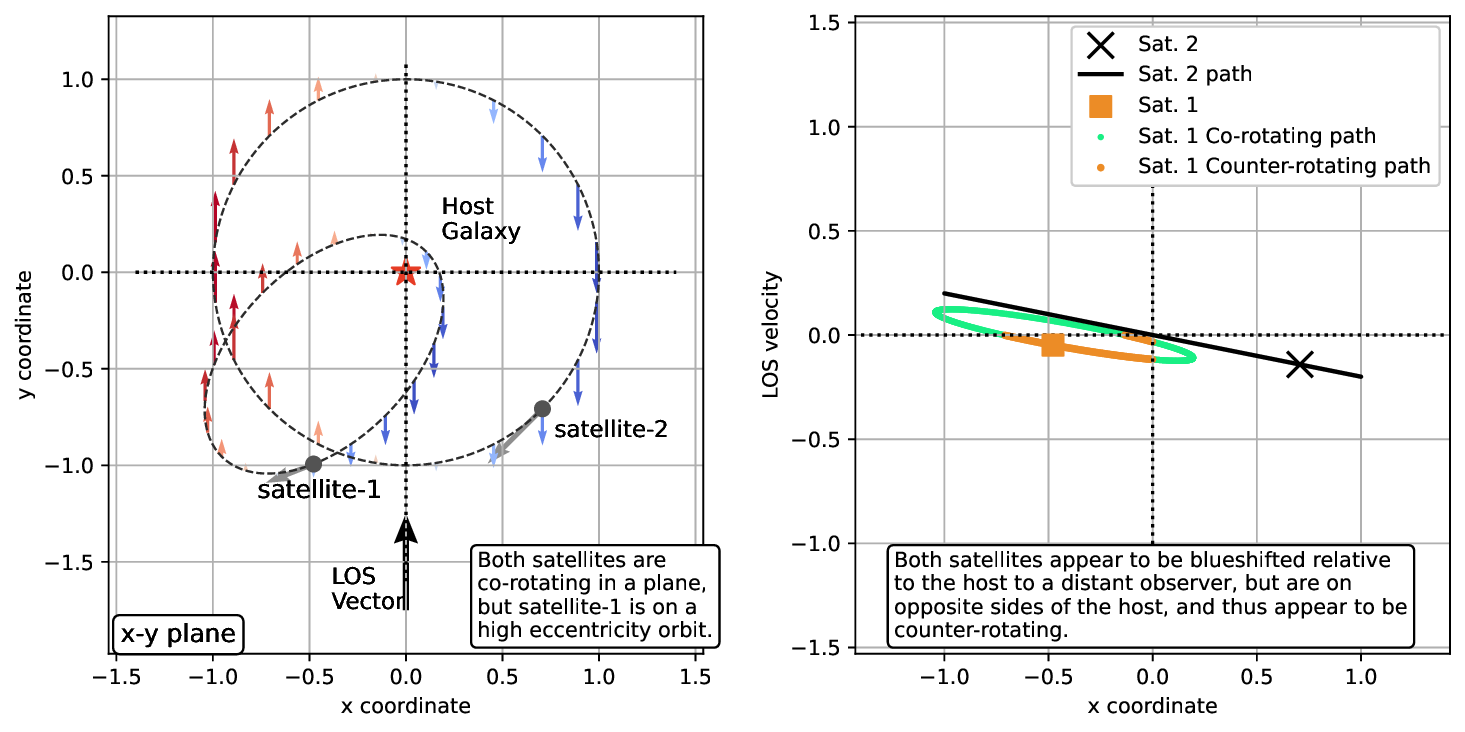}
	\caption{A diagram detailing how co-rotating in-plane satellites can appear to be counter-rotating to distant observers, if one of the satellites is on a highly eccentric orbit. The left panel demonstrates this observing the plane 'face-on', with coloured arrows at points along the orbital paths indicating the magnitude of the LOS velocity of the observer. In the right panel, the full path of satellite 1 and satellite 2 is plotted in the LOS velocity - x spatial coordinate space. The green path indicates the part of satellite 2's orbit where it appears to be co-rotating with satellite 1, while the orange part where its counter-rotating. Co-rotation can occur when $v_{\mathrm{los}}>0$ or $v_{\mathrm{los}}<0$ for both satellites, as long as both are on the same side of the host. Both satellites have $v_{\mathrm{los}}<0$, but reside on opposite sides of the host, and thus appear to be counter-rotating.}
	\label{fig:Ncorr_Ell_illustration}
\end{figure*}

\begin{figure}
	\centering
	\includegraphics[draft=false,width=7cm]{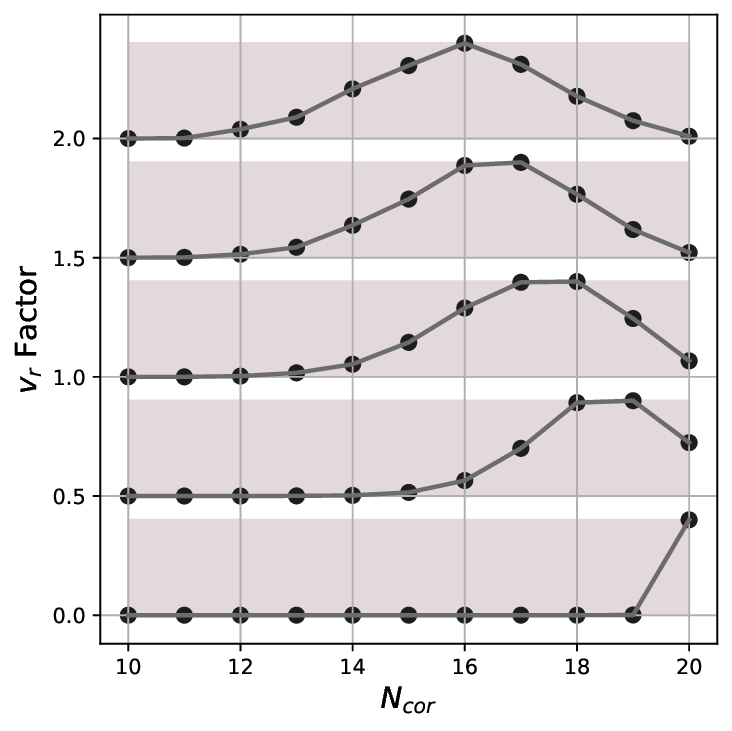}
	\caption{The distribution of $N_{\mathrm{cor}}$ built from 10,000 iterations of systems with edge-on, fully populated satellite planes ($N_{\mathrm{sat}}=20$, $f_{p}=1.0$, $\theta_{\mathrm{rot}}=0$) across multiplicative factors 0.0 to 2.0 for the $v_r$ velocity magnitude of simulated satellite orbits.}
	\label{fig:Ncor_Ell_KDE}
\end{figure}

One may notice that even in edge-on cases with fully populated satellites planes ($f_p=1.0$), that $N_{\mathrm{cor}}$ is not equal to the maximum possible value of 20. In that scenario, all satellites are co-rotating, so how can it be that it is often expected that not all satellites appear to be co-rotating to outside observers? When a satellite has an elliptical orbit, and the periapsis and apoapsis vectors are not parallel or anti-parallel to the viewing vector of a distant observer, it can appear to be counter-rotating compared to other satellites in the plane at certain phases of its orbit, just before or after it reaches periapsis and apoapsis. We illustrate this effect in Figure \ref{fig:Ncorr_Ell_illustration}. This typically occurs near apoapsis, where satellites on elliptical orbits are more likely to be found if a random orbital time is selected. Therefore the expected $N_{\mathrm{cor}}$ of fully populated satellite planes viewed edge-on is a function of the mean ellipticity of the orbits of satellites that comprise that plane. In Figure \ref{fig:Ncor_Ell_KDE}, we observe 10,000 iterations of fully populated ($f_p=1.0$) edge-on ($\theta_{\mathrm{rot}}=0^\circ$), but we multiply the $v_r$ velocity magnitudes of the satellites by a factor from 0.0 to 2.0. At 0.0 all orbits become circular, and at 2.0 they are effectively twice as elliptical. The figure demonstrates that in the case of purely circular orbits, the number of co-rotating satellites is always 20, the maximum value. But with increasing ellipticity in the orbits, the expected $N_{\mathrm{cor}}$ decreases.

\subsubsection{Detectability of Satellite Planes}

The metrics $N_{\mathrm{cor}}$, $b/a$, and $v_{\mathrm{los}}$ are commonly used to infer the presence of co-rotating, plane-like distributions of satellite galaxies. As demonstrated in the previous section, each of these metrics exhibits sensitivity to different aspects of planar configurations—specifically, to those arrangements which they most effectively distinguish from random or isotropic distributions. For instance, the axis ratio $b/a$ is particularly sensitive to edge-on planes (i.e., viewing angle $\theta_{\mathrm{rot}} = 0^\circ$), while the line-of-sight velocity metric $v_{\mathrm{los}}$ is more sensitive to face-on planes (i.e., $\theta_{\mathrm{rot}} = 90^\circ$).

In the broader context of satellite plane research, the precise values of these metrics are not typically the primary focus. Rather, they serve as statistical tools for hypothesis testing, addressing questions such as: Is this observed satellite configuration more disk-like than those produced in $\Lambda$CDM simulations? The observed values are compared to distributions derived from isotropic satellite configurations or from analogous environments within $\Lambda$CDM-based simulations. The degree to which the observed system deviates from these reference distributions—i.e., how much of an outlier it is—is what is usually reported.

In this section, we evaluate the utility of each metric in the context of such hypothesis testing.

We begin by defining a baseline distribution for comparison. For simplicity, we adopt an isotropic satellite distribution as the base case. We generate 30,000 mock observations with a plane fraction $f_p = 0$, and measure $N_{\mathrm{cor}}$, $b/a$, and $v_{\mathrm{los}}$ for each realisation. This yields baseline distributions for all three metrics under purely isotropic conditions.

Next, for a given plane configuration defined by specific values of $f_p$ and viewing angle $\theta_{\mathrm{rot}}$, we generate a mock observation and measure $N_{\mathrm{cor}}$, $b/a$, and $v_{\mathrm{los}}$. Each observed value is then compared to its corresponding isotropic baseline distribution. A "success" is recorded if the value is more extreme (a two-sided test) than 97.7\% of the baseline values, approximately corresponding to a 2$\sigma$ outlier in a Gaussian distribution. Therefore a baseline success rate of $2.3\%$ is expected as a result of a random chance alignments. For these mock observations, there are no simulated limitations such as limited sky-coverage or velocity uncertainty. We discuss the implications of this assumption in Section \ref{sec:discussion}.

\begin{figure*}
	\centering
	\includegraphics[draft=false,width=18cm]{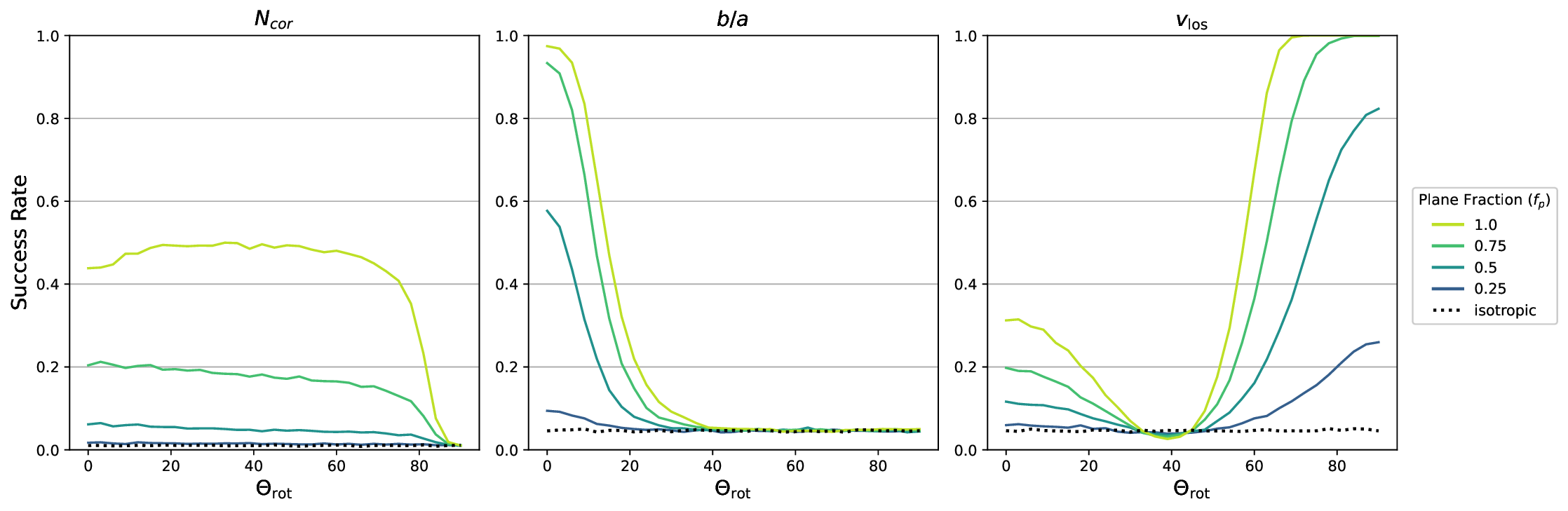}
	\caption{The "Success rate", or the rate at which a 2-sigma outlier is registered for a plane configuration compared to baseline isotropic distributions, using metrics $N_{\mathrm{cor}}$, $b/a$ and $v_{\mathrm{los}}$ across plane fraction ($f_p$), represented with coloured lines and a legend in the figure, and the plane inclination ($\theta_{\mathrm{rot}}$) on the x-axis. }
	\label{fig:Pfrac_Rot_Success_Rate}
\end{figure*}

To evaluate the consistency of detection, we repeat this process across 10,000 mock observations for each $(f_p, \theta_{\mathrm{rot}})$ configuration. The success rate is defined as the fraction of these observations that meet the 2$\sigma$ outlier criterion. We report the success rate as a function of $f_p$ and $\theta_{\mathrm{rot}}$ for a fixed number of 20 satellites in Figure \ref{fig:Pfrac_Rot_Success_Rate}.

Figure \ref{fig:Pfrac_Rot_Success_Rate} matches expectations established in Figure \ref{fig:simple_KDE}. For $N_{\mathrm{cor}}$, the success rate is flat across all inclination states except the face-on scenario. For $b/a$ it is high when a plane is viewed edge-on and low otherwise. For $v_{\mathrm{los}}$, it is high when a plane is viewed face-on and low otherwise. In between these extremes, this figure highlights the low success rates of identifying non-isotropic distributions with low plane fraction ($f_p<0.5$) and at intermediate inclination ($\theta_{\mathrm{rot}}=20-60^\circ$).

\subsubsection{Observational Constraints}

To support the goals of this study, we aim to translate the success rate framework into a set of quantitative guidelines for observational programs targeting satellite planes. In particular, a key consideration for observers is determining the minimum $N_{\mathrm{sat}}$ required to achieve statistically significant detections. This sets practical constraints on which host environments are viable survey targets. Specifically, those with a sufficient $N_{\mathrm{sat}}$ above the luminosity threshold, which itself depends on the instrument sensitivity and configuration \footnote{This luminosity threshold often hinges on the capabilities of the spectrograph used to measure the LOS velocities, and is typically $\mu_{o,g}\sim24.5\ \mathrm{mag}/\mathrm{arcsec}^2$ for instruments such as MUSE}.

\begin{figure*}
	\centering
	\includegraphics[draft=false,width=18cm]{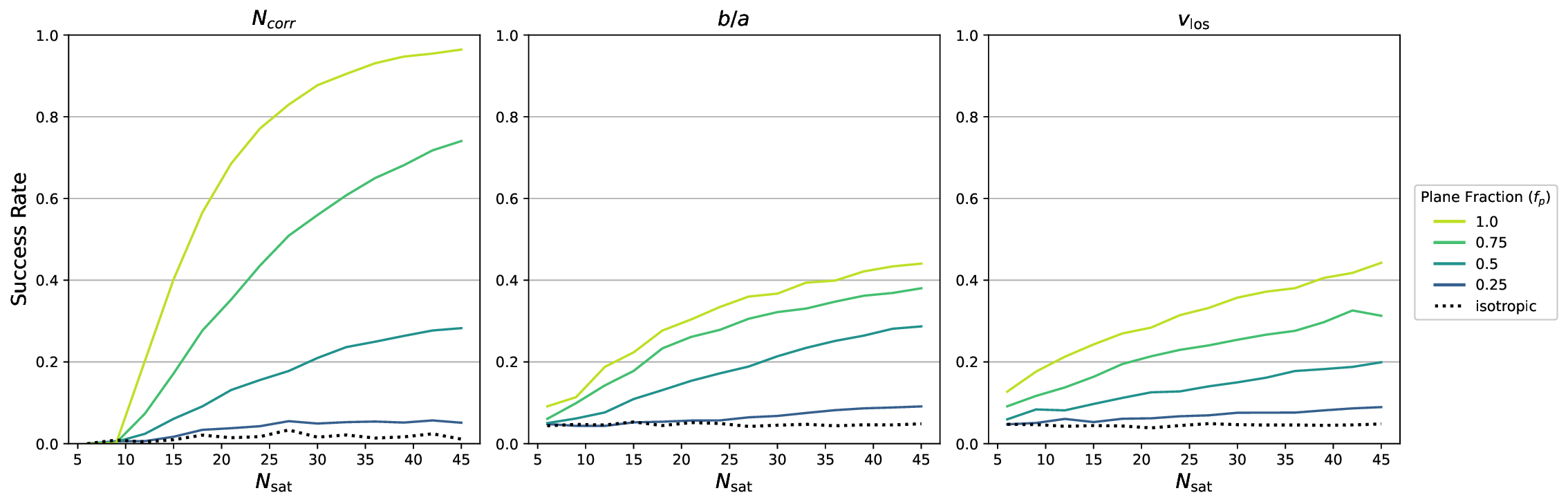}
	\caption{The "Success rate", or the rate at which a mock observation of $N_{\mathrm{cor}}$, $b/a$ or $v_{\mathrm{los}}$ is a 2-sigma outlier compared to the isotropic case across plane fraction ($f_p$, coloured lines as annotated in the figure) with randomised inclination ($\theta_{\mathrm{rot}}$) for varying numbers of satellites ($N_{\mathrm{sat}}$, x-axis).}
	\label{fig:Pfrac_N_Success_Rate}
\end{figure*}

To explore these observational constraints, we repeat the previous experiment, this time drawing the plane orientation randomly from a uniform distribution over the unit sphere in each iteration. We evaluate the success rate as a function of both $f_p$ and the number of satellites in the system ($N_{\mathrm{sat}}$). Since orientations are sampled isotropically, this setup naturally favors edge-on planes over face-on ones, as there are geometrically more ways to observe a system at low inclination. Specifically, the values of $\theta_{\mathrm{rot}}$ are described by a PDF with the function $f(x)=\mathrm{cos}(x)$. The resulting success rates are shown in Figure \ref{fig:Pfrac_N_Success_Rate}.

This figure indicates that, even when surveying systems with large numbers of satellites, the overall success rate remains below 50\% if host systems are sampled randomly. This is largely driven by the frequency at which planes are observed at inclinations that our chosen metrics are not sensitive to, which based on Figure \ref{fig:Pfrac_Rot_Success_Rate}, is about 20 to 60 degrees, which accounts for 50\% of orientations in an isotropic distribution on a unit sphere. Notably, the success rate for the $N_{\mathrm{cor}}$ metric remains low across most configurations, except in the case of fully populated planes ($f_p = 1.0$), due to the discrete nature of the statistic.

\subsubsection{Informed Observing Strategies}

However, observers searching for satellite planes may choose a more informed strategy for selecting targets, rather than relying on purely random sampling. For instance, a preliminary assessment of a system’s axis ratio ($b/a$) could be used to estimate its inclination angle $\theta_{\mathrm{rot}}$, thereby identifying environments more likely to host detectable plane-like structures. Such pre-selection introduces a constraint on $\theta_{\mathrm{rot}}$, which can be exploited to improve the chances of identifying statistically significant signatures.

\begin{figure*}
	\centering
	\includegraphics[draft=false,width=18cm]{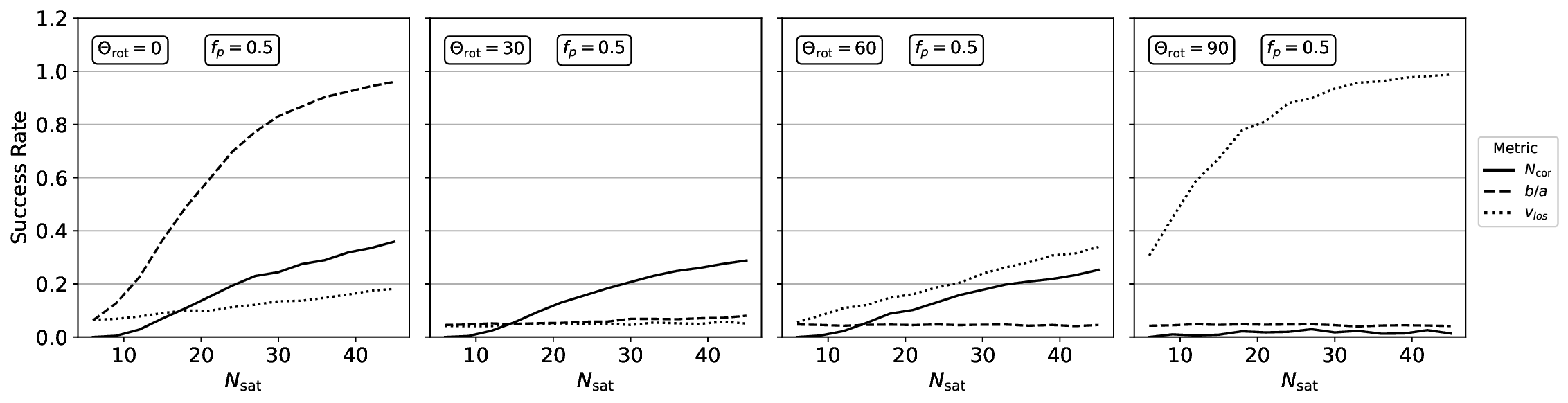}
	\caption{The "Success rate", or the rate at which a mock observation of $N_{\mathrm{cor}}$, $b/a$ or $v_{\mathrm{los}}$ (different lines as annotated in the figure) is a 2-sigma outlier compared to the isotropic case across varying number of satellites ($N_{\mathrm{sat}}$, x-axis). $f_p$ is fixed at $0.5$ and $\theta_{\mathrm{rot}}$ varies as shown in each panel.}
	\label{fig:Pfrac_N_fixed_rot_Success_Rates}
\end{figure*}

To support this approach, we provide a grid of success rates as a function of the $N_{\mathrm{sat}}$ for selected fixed values of $\theta_{\mathrm{rot}}$ in Figure~\ref{fig:Pfrac_N_fixed_rot_Success_Rates}. If observers adopt a reasonable filtering strategy, such as prioritising systems likely to be oriented face-on ($\theta_{\mathrm{rot}} = 90^\circ$) by excluding flattened systems with low $b/a$, the probability of detection could increase. For example, success rates based on the $v_{\mathrm{los}}$ metric exceeds 80\% for $N_{\mathrm{sat}}=20$ even for moderately populated satellite planes, as demonstrated in Figure \ref{fig:Pfrac_N_fixed_rot_Success_Rates}. This suggests that even modest pre-selection strategies, when guided by geometric considerations, can enhance the scientific return of satellite plane surveys.

\subsubsection{Multivariate Comparisons}

The preceding figures evaluate success rates for each metric in isolation. However, in practice, satellite plane analyses often consider multiple metrics simultaneously to assess whether an observed system is a joint outlier. 

We assess the multivariate extremeness of each observation in the three-dimensional parameter space defined by $N_{\mathrm{cor}}$, b/a, and $v_{\mathrm{los}}$, using a multivariate normal distribution (MVN, \textsc{multivariate\_normal} from \textsc{Scipy} \citep{2020SciPy-NMeth}). The exact process is implemented as follows:
\begin{enumerate}
    \item We generate 30,000 isotropic systems with $N_{\mathrm{sat}}=20$, and measure $N_{\mathrm{cor}}$, b/a, and $v_{\mathrm{los}}$ in each system from a mock-observation.
    \item We fit a MVN distribution to the parameter space of these three metrics.
    \item We sample 10,000 points (consisting of $N_{\mathrm{cor}}$, b/a, and $v_{\mathrm{los}}$) from this distribution and calculate the distribution p-values for them.
    \item This forms a one-dimensional p-value distribution for the isotropic baseline case.
    \item Generate 10,000 systems of satellites for each combination of $f_p$ with random $\theta_{\mathrm{rot}}$, measure $N_{\mathrm{cor}}$, b/a, and $v_{\mathrm{los}}$, determine p-value of this observation from the isotropic MVN distribution.
    \item If this p-value lies in the bottom $2.3\%$ of the isotropic p-value distribution, a success is registered.
\end{enumerate}
Following the same practice established before, the 'success rate' describes the rate at which 2-sigma outliers of the MVN distribution formed by isotropic base case are registered for a satellite plane.

\begin{figure}
	\centering
	\includegraphics[draft=false,width=8cm]{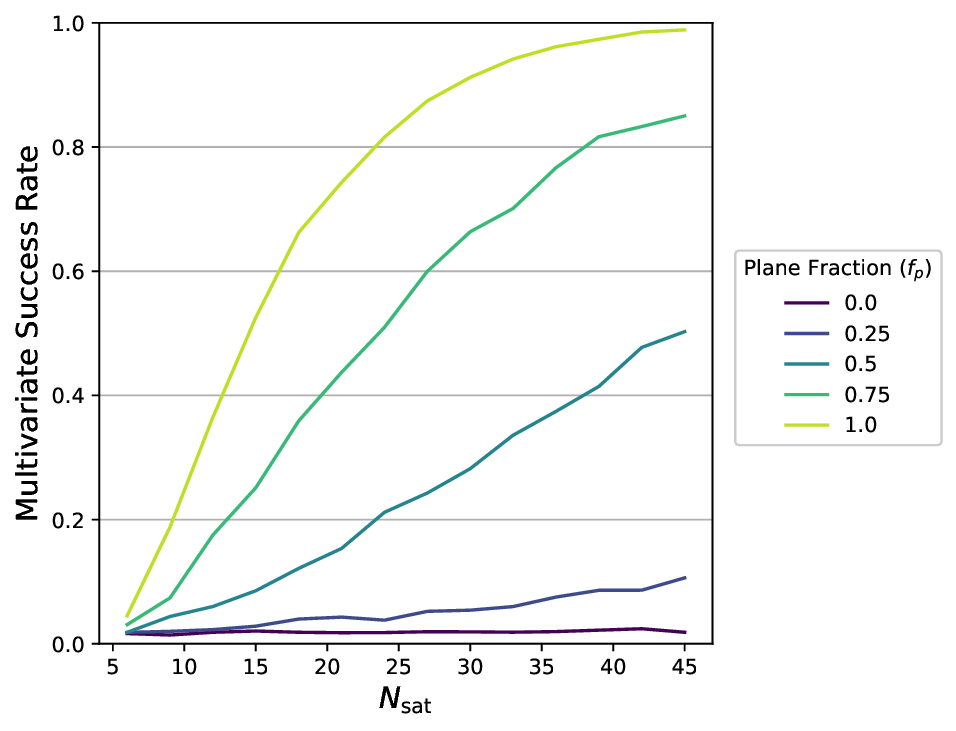}
	\caption{The Multivariate "Success rate", or the rate at which a 2-sigma outlier of in the $N_{\mathrm{cor}}$, $b/a$ or $v_{\mathrm{los}}$ three-dimensional parameter space is measured for plane configurations across plane fraction ($f_p$, represented with distinct lines in the figure) and the number of satellites ($N_{\mathrm{sat}}$), with randomised inclination ($\theta_{\mathrm{rot}}$).}
	\label{fig:TOTAL_Success_Rate}
\end{figure}
\begin{figure*}
	\centering
	\includegraphics[draft=false,width=18cm]{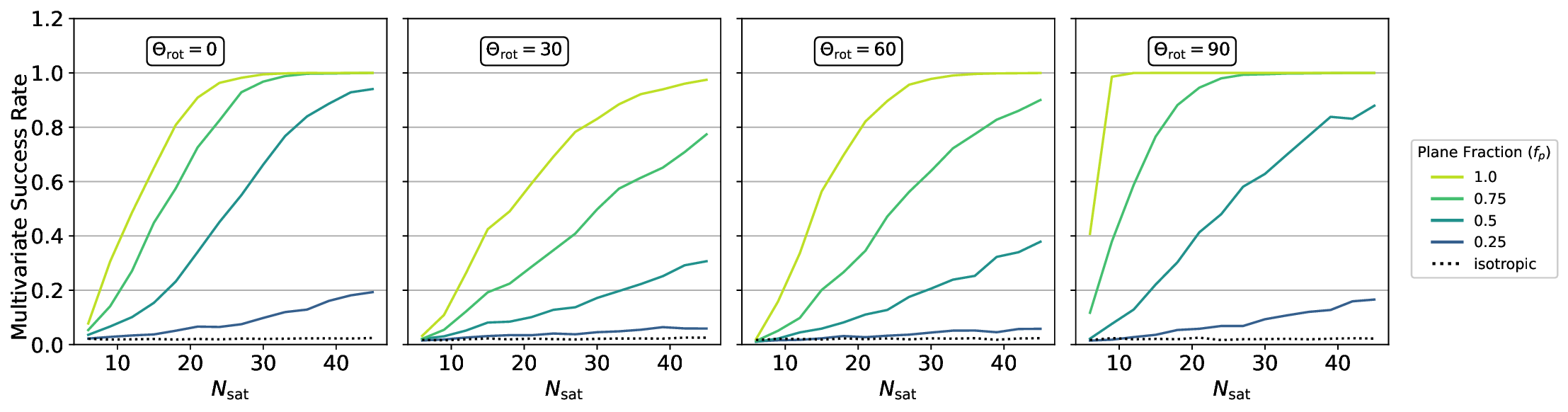}
	\caption{The Multivariate "Success rate", or the rate at which a 2-sigma outlier of in the $N_{\mathrm{cor}}$, $b/a$ or $v_{\mathrm{los}}$ three-dimensional parameter space is measured for plane configurations across plane fraction ($f_p$, represented with distinct lines in the figure) and the number of satellites ($N_{\mathrm{sat}}$), with fixed values of inclination ($\theta_{\mathrm{rot}}$) as annotated in the figure.}
	\label{fig:TOTAL_fixed_rot_Success_Rate}
\end{figure*}

In Figure \ref{fig:TOTAL_Success_Rate}, we present the multivariate success rate as a function of $f_p$ and $N_{\mathrm{sat}}$, with randomly oriented planes. Similarly, Figure \ref{fig:TOTAL_fixed_rot_Success_Rate} shows success rates for fixed orientations, allowing for comparison under targeted observational strategies.

When considering these multivariate success rates, we find that combining all three metrics has a complex influence on the performance. Comparing Figures \ref{fig:Pfrac_N_Success_Rate} and \ref{fig:TOTAL_Success_Rate}, we find that at $N_{\mathrm{sat}}=20$ and $f_p=0.5$ with random rotation, the success rate improves from $\sim12\%$ (while using $N_{\mathrm{cor}}$ in isolation, as in Figure \ref{fig:Pfrac_N_Success_Rate}) to $\sim15\%$ in the multivariate case (as in Figure \ref{fig:TOTAL_Success_Rate}) suggesting a modest performance improvement. But in other cases the performance is reduced, as seen when comparing Figures \ref{fig:Pfrac_N_fixed_rot_Success_Rates} and \ref{fig:TOTAL_fixed_rot_Success_Rate} for the $f_p=0.5$ case. In that instance, combining the metrics using MVN degrades the performance compared to using the $v_{\mathrm{los}}$ metric alone specifically for the face-on plane case. The MVN method is robust because it makes no assumption about the orientation of a satellite plane \textit{a-priori}. But if the orientation of a satellite plane can be constrained to either the face-on or edge-on case (ie. $\theta_{\mathrm{rot}}$ can be estimated \textit{a-priori}), then this result suggests a more focused detection strategy will be more effective. We highlight two mechanisms that explain this phenomenon.

First, each metric exhibits sensitivity to different aspects of satellite plane geometry. Incorporating a metric poorly suited to detecting planes of a given configuration contributes little additional discriminative power and may even dilute the overall signal. 

Second, the metrics are not independent. For instance, in our simulations of thin, co-rotating planes, systems observed to have small $b/a$ (i.e., thin planes) are often viewed edge-on, which naturally corresponds to high $N_{\mathrm{cor}}$ and high $v_{\mathrm{los}}$. In such cases, the joint detection of both a low $b/a$ and a high $N_{\mathrm{cor}}$ does not provide significantly more information than either metric alone, as the likelihood of an isotropic system mimicking both features simultaneously is already vanishingly small.

This behaviour arises from our simplified example. We specifically restrict the host environments to isolated hosts. We compare only isotropic and flattened, co-rotating disks. But outside of simulations, it is more complex than this simple dichotomy. Systems of satellites involved in merger events or galactic interactions can display characteristics not accounted for in our study. For example, recent studies show that two $L_*$ galaxies in the process of merging can produce strong co-rotation signals in their satellite distributions \citep{Pawlowski_2024,Jerjen2025}. However, these 'co-rotating mergers' are not expected to produce stable satellite planes \citep{Kanehisa2023}. This class of system may frequently show strong co-rotation, but not a thin spatial disk. It is these kinds of systems that may be effectively filtered using a multi-variate approach.

Interestingly, the highest overall success rates are achieved for face-on planes ($\theta_{\mathrm{rot}} = 90^\circ$), driven primarily by the $v_{\mathrm{los}}$ metric. This is likely due to reduced observational complexity: edge-on planes can suffer from ambiguous kinematic signatures, such as apparent counter-rotation in elliptical orbits, and are more susceptible to outliers in the $b/a$ metric. In contrast, face-on planes allow for cleaner line-of-sight velocity signals and less ambiguity in plane structure, leading to more robust detections.

Unfortunately, face-on planes are intrinsically rare. From Figure \ref{fig:Pfrac_Rot_Success_Rate}, we find that the highest success rates for $v_{\mathrm{los}}$ occur at inclinations of $60 - 90^{\circ}$, which represent only about $\sim13\%$ of all orientations under a uniform distribution on the unit sphere. Similarly, $b/a$ is most effective for detecting edge-on planes ($\theta_{\mathrm{rot}} = 0 - 20^{\circ}$), which make up about $\sim34\%$ of cases. These geometric constraints fundamentally limit the detection efficiency of even the best-performing metrics in realistic survey conditions.

\section{Discussion} \label{sec:discussion}

In our analysis, we have assumed idealised conditions, neglecting observational limitations or uncertainties. In practice, there are several sources of uncertainty that could influence the detection and characterisation of satellite planes. Here, we identify four key observational constraints: limited sky coverage, line-of-sight velocity uncertainty, host galaxy mass uncertainty and contamination.

For the former, if a survey targets a nearby or massive host galaxy, the virial radius may subtend a large angle on the sky, potentially larger than the survey’s field of view, as is the case for the MATLAS \citep{Heesters2021} for example. In such cases, satellites near the outskirts may fall outside the observed region, effectively reducing the apparent satellite count. This is particularly important for the $b/a$ metric, which is sensitive to the spatial distribution and especially influenced by outliers at large projected radii. However, with the advent of deep, wide-field surveys such as the DESI Legacy Imaging Surveys \citep{DECALS,Zou_2017,Zou_2018}, this issue is becoming progressively less limiting.

Line-of-sight velocity uncertainty introduces a second important observational effect. While emission-line dwarf galaxies typically have precise velocity measurements (of 1–5 $\mathrm{km}\,\mathrm{s}^{-1}$, depending on spectral resolution \citep{Muller2018,Karachentsev2024,Crosby2024}), most satellite galaxies are quenched, early-type systems. These rely on absorption line measurements, which often have higher uncertainties (10–30 $\mathrm{km}\,\mathrm{s}^{-1}$ \cite{Muller2018,Crosby2024}). For edge-on planes where relative satellite velocities are large (100–300 $\mathrm{km}\,\mathrm{s}^{-1}$ in Milky Way-like systems), uncertainties of this amplitude are unlikely to have a significant impact.

However, $v_{\mathrm{los}}$ is most sensitive to face-on planes. In these configurations, in-plane satellites have small relative projected velocities—typically $v_{\mathrm{los}} \sim 0.1$ (as in Figure \ref{fig:simple_KDE}), which for Milky Way analogs with mass $M_{\mathrm{vir}}\sim1.5\times10^{12}\,M_{\odot}$ \citep{McMillan2016} and thus virial velocity $v_{\mathrm{vir}}\sim150\,\mathrm{km}\,\mathrm{s}^{-1}$, is about 15 $\mathrm{km}\,\mathrm{s}^{-1}$. Since this value is comparable to the measurement uncertainty of early-type dwarfs, some scatter is expected. Nevertheless, the difference in expected $v_{\mathrm{los}}$ between face-on planes and isotropic systems is large: for $f_p=0.5$, (as in Figure \ref{fig:simple_KDE}) this difference is $v_{\mathrm{los}} \sim 0.3$, or about 45 $\mathrm{km}\,\mathrm{s}^{-1}$ for Milky Way analogs. For massive Cen A / M104 analogs with mass $M_{\mathrm{vir}}\sim1\times10^{13}\,M_{\odot}$ \citep{Karachentsev2020_b,Muller2022}, the expected $v_{\mathrm{los}}$ difference is about 90 $\mathrm{km}\,\mathrm{s}^{-1}$. As a result, velocity uncertainty is unlikely to erase the signal entirely, though it may reduce contrast between plane configurations in the most favorable cases of lower-mass systems.

The virial mass is used to calculate the virial velocity, the normalising factor in $v_{\mathrm{los}}$. To assess differences between expected and observed LOS velocities and infer the possible presence of a satellite plane, a measurement of the virial mass independent of the motions of the satellites is required. We previously established that the expected velocity contrast for face-on and isotropic systems is $\sim45\,\mathrm{km}\,\mathrm{s}^{-1}$ for Milky Way–type halos with virial mass $1.5\times10^{12}\,M_{\odot}$. If we set the virial velocity uncertainty to $45\,\mathrm{km}\,\mathrm{s}^{-1}$ and propagate the uncertainty to the virial mass, this implies the mass and its uncertainty would be $1.5^{+1.7}_{-0.9}\times10^{12}\,M_{\odot}$. The uncertainty of the mass measurement should be no larger than that in this result in order to differentiate face-on with isoptropic planes. High precision measurements of the Milky Way's virial mass are about $1.3^{+0.3}_{-0.3}\times10^{12}\,M_{\odot}$ \citep{McMillan2016}, so falling within this uncertainty limit is possible when high-precision mass estimates are available. For more distant galaxies where high-precision mass measurements aren't available, then abundance matching against the host's stellar mass \citep{Wechsler2018,Stiskalek2021,Crosby_2023_b} is frequently a method of choice. This involves multiplying the host's stellar mass by the total to stellar mass ratio for that class of galaxy. This ratio can be established from observational programs \citep{Heymans_2006}, which typically have uncertainty of $\pm30\%$, or established from cosmological simulations which largely agree with observational findings \citep{Crosby_2023_b}. Provided that the stellar mass uncertainty is small ($\sim\pm10\%$), then $\pm30\%$ mass uncertainty is sufficient to achieve a precise enough mass estimate.

In our mock observations, we deliberately do not include any lop-sidedness in the satellite distributions. This is the phenomenon where satellite orbital phases appear to spatially clustered \citep{Libeskind2016,Pawlowski2017,Heesters2024}, even around isolated host galaxies \citep{Brainerd2020,Wang2021}. The degree of lop-sidedness is dependent on host galaxy morphology and the distribution of satellites \citep{Brainerd2020,Heesters2024}. Therefore observations of lop-sided systems must be considered carefully when compared to our results.

We also exclude galaxies which the simulation has identified as having a non-cosmological origin from our selection of \textsc{TNG100} systems (\textsc{SubhaloFlag} set to 0). These are dark matter deficient objects, which may be fragments from baryonic processes internal to the host galaxy, or otherwise disrupted satellite galaxies from interactions. The simulation cannot clearly differentiate between these two cases and thus it is recommended to exclude these objects \citep{Nelson_2019}. However, carefully including them across different cosmological contexts may be important in properly reproducing phenomena involving galaxies on the low-mass end of the mass function \citep{Guo2011,SantosSantos2025}—including satellite planes \citep{Sawala2022}. For example, galaxies of non-cosmological origin include dark matter deficient tidal dwarf galaxies, whose formation has been identified as a possible mechanism for forming satellite planes \citep{Kroupa_2005,Metz2007_b,Pawlowski2012_a,Hammer_2013}. We adopt the conservative approach of removing all objects of non-cosmological origin, and therefore dark matter deficient tidal dwarf galaxies, to generate radial distance and velocity distributions that aren't contaminated by baryonic fragments created in internal galaxy processes.


This study also underscores the importance of targeting rich satellite systems, with $N_{\mathrm{sat}} \sim 10$ emerging as a practical limit where the rate of successfully detecting a planar arrangement of satellites is very low, around $\sim5\%$ for moderately populated planes ($f_p=0.5$) as in Figure \ref{fig:TOTAL_fixed_rot_Success_Rate}. However, designing surveys around this threshold introduces potential selection biases that must be carefully considered.

Most notably, spectroscopic limits—typically around $m_g \lesssim 18$ \citep{Muller2021,Karachentsev2024,Crosby2024} exclude the faintest satellites, meaning that systems with $\geq$10 detectable satellites are inherently rare. In practice, many such systems reside in galaxy clusters, massive galaxy groups, or interacting galaxy pairs, rather than isolated Milky Way analogs. This raises a key concern: the kinematic and spatial configurations of these high-mass environments may differ significantly from the well-studied satellite systems of Milky Way-like hosts.

As a result, conclusions drawn from such rich systems may not be representative of more typical satellite environments. Whether planar structures in these massive systems follow the same formation channels or even exhibit similar statistical properties, remains an open question, and a potential source of bias in interpreting observational results.

\begin{figure}
	\centering
	\includegraphics[draft=false,width=7cm]{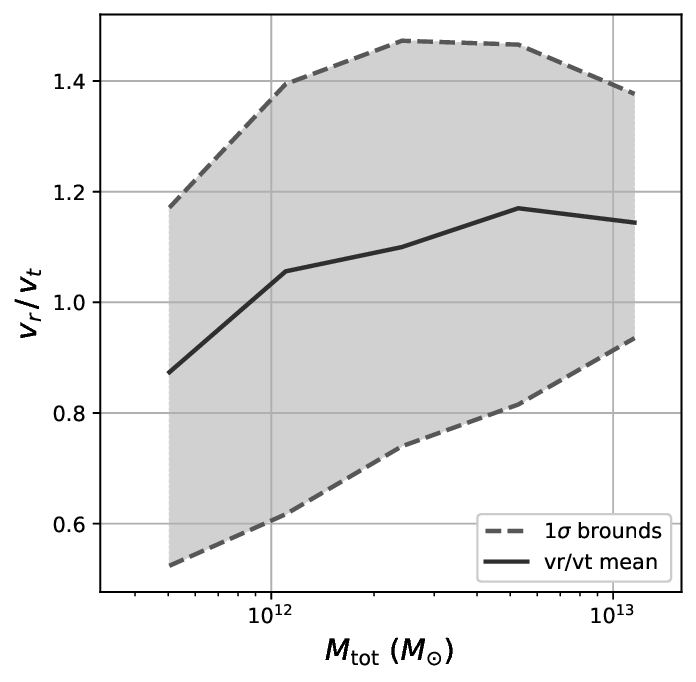}
	\caption{The mean radial and tangential velocity ratio ($v_r/v_t$) with 1$\sigma$ uncertainty bounds of satellites as a function of host total mass ($M_{\mathrm{tot}}$) of the 920 \textsc{TNG100} host systems used in this paper. The solid line represents the mean, and the dotted lines the 1$\sigma$ uncertainty bound.}
	\label{fig:vrvt_mass_corr}
\end{figure}

Though we have attempted to eliminate the influence of the mass of the host galaxy by normalising our measured quantities by the virial radius and virial velocity, there still remain slight correlations between the orbital properties of satellites and the host mass. The most significant of these correlations lies between the ellipticity of satellite galaxy orbits and the host galaxy mass. In Figure \ref{fig:vrvt_mass_corr} we visualise the trend in the radial / tangential velocity ratio ($v_r/v_t$) of satellites as a function of host total mass. We discover that satellite orbits become more elliptical as host mass increases. As shown in Figures \ref{fig:Ncorr_Ell_illustration} and \ref{fig:Ncor_Ell_KDE}, this can degrade the effectiveness of the $N_{\mathrm{cor}}$ metric. The $N_{\mathrm{cor}}$ metric therefore may not be successful in distinguishing the presence of satellite planes around high-mass hosts.

Finally, our simulations here do not include contamination. In this context, contamination refers to dwarf galaxies that are incorrectly assigned as satellites of $L_*$ galaxies that are nearby in the on-sky projection, but may be separated by many Mpc along the line of sight. For example, nearby galaxy groups or clusters in the plane of sky may have high peculiar velocities that make it difficult to properly identify the system of dwarf galaxies in that region of the sky based on LOS velocities alone. The extent of contamination varies, and must be assessed carefully on a case-by-case basis in observing programs. Ideally, targets of observing programs should be not only isolated spatially, but also in LOS velocity space.

\section{Summary and Conclusion}

We build a simple model to generate mock observations of the spatial and kinematic distributions of satellite galaxies around isolated hosts. This model is used to assess how effectively various observable metrics can determine the presence of satellite planes within these systems. We sample these satellite distributions from comparable host environments in the Illustris TNG100 simulation. From these, we generate mock observations simulating distant observers who measure the on-sky positions and line-of-sight velocities of the satellites. We adjust the spatial and kinematic distributions of satellites to control their membership in a thin, co-rotating satellite plane. We examine how effectively three metrics—$N_{\mathrm{cor}}$ (number of co-rotating satellites), $b/a$ (on-sky flattening), and $v_{\mathrm{los}}$ (mean absolute line-of-sight velocity)—diagnose the presence of planar structure in satellite systems. We make this assessment across the fraction of satellites inside the plane ($f_p$), the inclination of the satellite plane relative to the observer ($\theta_{\mathrm{rot}}$) and the number of satellites within the system ($N_{\mathrm{sat}}$). In particular, we define a “success rate” as the frequency with which these metrics register as $2\sigma$ outliers from the isotropic distribution, indicating a statistically significant detection of planarity.

\begin{itemize}
    \item The success rate for $N_{\mathrm{cor}}$ is moderate ($20-50\%$) for highly populated satellite planes ($f_p\geq0.75$) across all inclinations, except for face-on satellite planes where the success rate is near zero.
    \item This success rate for $N_{\mathrm{cor}}$ is comparably low even in the most favourable cases because in-plane satellites on highly eccentric orbits can appear to be counter-rotating to distant observers.
    \item The success rate for $b/a$ exceeds 50\% for moderately populated planes ($f_p\geq0.5$), but only when the plane is nearly edge-on ($\theta_{\mathrm{rot}}<10^\circ$); in all other orientations, the success rate is near zero.
    \item The success rate for $v_{\mathrm{los}}$ is high ($>50\%$) for moderately populated planes ($f_p\geq0.5$), but only for face-on planes ($\theta_{\mathrm{rot}}\sim60-90^\circ$), otherwise it has a low success rate ($<20\%$).
    \item Detecting any plane with success rate $>50\%$ requires at least 20 satellites in either an edge-on, or face-on plane configuration.
    \item For all satellite planes that are not completely populated ($f_p<1.0$), the success rate for systems with less than 20 satellites ($N_{\mathrm{sat}}<10$) is always low ($\sim5\%$), except for face-on satellite planes.
    \item Satellite planes with low population fractions ($f_p\leq0.25$) and moderate inclinations ($\theta_{\mathrm{rot}}\sim20\text{-}60^\circ$) are difficult to distinguish from isotropic distributions unless the system contains a very large number of satellites ($N_{\mathrm{sat}}>40$).
    \item Combining these metrics together into a three dimensional parameter space to assess the success rate collectively offers modest performance improvements when the orientation is not known \textit{a-priori}. If the orientation can be constrained \textit{a-priori}, particularly to face-on or edge-on cases, focused plane detection strategies are likely to be much more effective.
    \item Face-on satellite planes have the highest success rate through the $v_{\mathrm{los}}$ metric, but such planes account for only 13\% of possible plane inclinations. This method relies upon a precise measurement of the host halo mass within $\pm30\%$.
    \item Edge-on satellite planes with high success rate through the $b/a$ metric account for 34\% of possible plane inclinations, about 3 times as many face-on planes.
\end{itemize}

These constraints are intended to provide an initial evidence base for the design of observing programs that aim to quantify the presence or absence of satellite planes in nearby galaxy groups. With the increasing over-subscription rates of 8m class optical IFU spectrographs (in particular MUSE), careful and well justified observing proposals are becoming yet more essential in receiving acceptance for those observing programs.

\section*{Acknowledgements}

O.M. is grateful to the Swiss National Science Foundation for financial support under the grant number PZ00P2\_202104.

M.S.P. acknowledges funding via a Leibniz-Junior Research Group (project number J94/2020)


\section*{Data Availability}
The data underlying this article will be shared on reasonable request to the corresponding author.



\bibliographystyle{mnras}
\bibliography{bibliography.bib} 


\clearpage 
\appendix

\bsp	
\label{lastpage}
\end{document}